# Performance study of a directional fast neutron detection system based on TPC and plastic scintillation detectors

Yidong Fu,[a,b] Yang Tian,[a,b,*] Yulan Li,[a,b] Yigang Yang,[a,b] Yiming Cai,[a,b] Jian Yang,[a,b] and Jin Li[a,b]

[a] *Department. of Engineering Physics, Tsinghua University,*
 *Beijing 100084, China*

[b] *Key Laboratory of Particle & Radiation Imaging (Tsinghua University), Ministry of Education,*
 *Beijing 100084, China*
 *E-mail*: `yangt@mail.tsinghua.edu.cn`

ABSTRACT: A neutron time projection chamber can locate the approximate direction of a neutron hot spot with high efficiency and a $4\pi$ field of view. The angular resolution can be significantly improved by adding several plastic scintillation detectors and using coincidence events. The specific performances of such a coincidence imaging system are studied based on theoretical calculations and experimental results. The calculated value of the angular resolution is approximately 2 °for the current system, which agrees well with the experimental results and sets an upper limit for the angular resolution of traditional back projection based online reconstruction methods. Although the statistical iterative method can breakthrough this limit and further improve the angular resolution, the time consumption is usually a problem. The coincidence imaging system can be further optimized for future applications based on the theoretical model.

---

[*]Corresponding author.

# Contents

# 1. Introduction

Special nuclear materials (SNMs) are nuclear materials containing $^{235}$U, $^{239}$Pu or other radioactive nuclides that can be used in nuclear weapons. SNMs emit gamma-rays and neutrons during active and passive interrogation. Some applications prefer to use fast neutrons to image SNMs because of their lower backgrounds and higher penetration through high-Z materials.[1] The direction of the SNM source can be determined using neutron scatter imaging systems. These systems can also be used in applications such as in the imaging of solar neutrons [2] and in thermo-nuclear fusion plasma diagnostics [3].

The neutron time projection chamber (TPC)[4][5] and the neutron scatter camera[6]-[8] are two types of neutron scatter imaging systems designed for fast neutron imaging in the past decade. The neutron TPC (Figure 1 (a)) is developed for fast neutron imaging with high efficiency and a $4\pi$ field of view (FOV). It can reconstruct the tracks of the recoil protons and determine the direction of the neutron source by averaging the directions of the recoil protons [4]. The neutron scatter camera (Figure 1(b)) images neutron sources with double scatter events. The direction of the incident neutron can be restricted to a conical surface to improve the angular resolution.

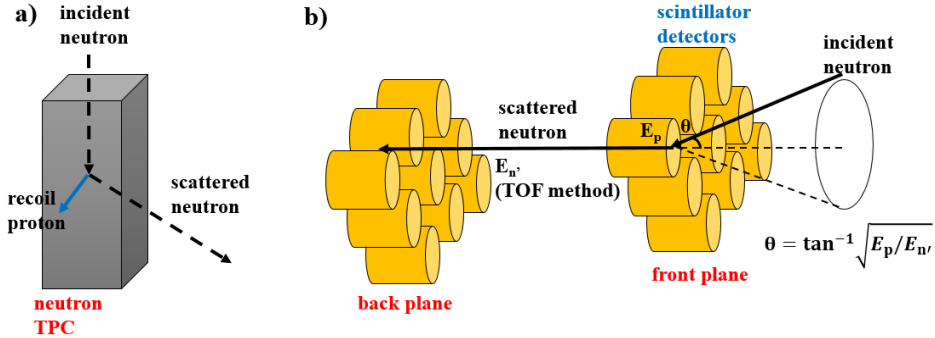

**Figure 1.** Directional fast neutron detectors: (a) Neutron TPC; (b) Neutron scatter camera

We proposed a method to improve the angular resolution of the neutron TPC by adding several plastic scintillation detectors and using the coincidence events [9] (Figure 2). Besides fast imaging of the hot spots with a $4\pi$ FOV using single scatter events in the TPC, the detector system can also provide high-resolution imaging using double scatter events: first in the TPC and subsequently in the scintillation detector. Dual-end readout scintillation detectors are used to provide the 3D position of the second scatter [10].

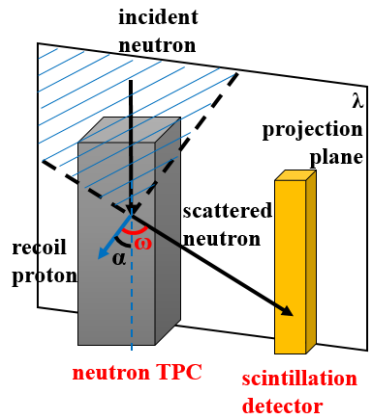

**Figure 2.** The imaging principle of the double scatter events.

The imaging principle of the double scatter events is shown in Figure 2. The first scatter occurs in the TPC, and then the scattered neutron is detected by the scintillation detector. The direction of the recoil proton is reconstructed by the TPC, and the direction of the scattered neutron is decided by the positions of the two scattering points. With the two directions measured, the direction of the incident neutron can be limited to a quarter of a plane (the dashed area in Figure 2). The incident direction can be further limited using the energy information. In an elastic scattering with a proton, the scattered neutron is perpendicular to that of the proton (angle $\omega$ is equal to 90°). [11] This fact can be used to distinguish effective events from the background and improve the angular resolution.

The details of the prototype system are presented in Section 2, including the detector design and the methods used in the data analysis. In Section 3, a theoretical model is proposed to evaluate the performances of the imaging system. The key factors affecting the angular resolution, efficiency, and FOV are discussed. In Section 4, the experimental results using different reconstruction algorithms are presented and compared with the theoretical limit. The coincidence imaging system can be further optimized based on the theoretical model for future applications.

## 2. Design of the prototype system

A prototype system is built to test the imaging principle of the double scatter events. The prototype system consists of a TPC and four plastic scintillation detectors. The scintillation detectors are placed 21 cm from the center of the TPC and separated by 45° from each other, as shown in Figure 3.

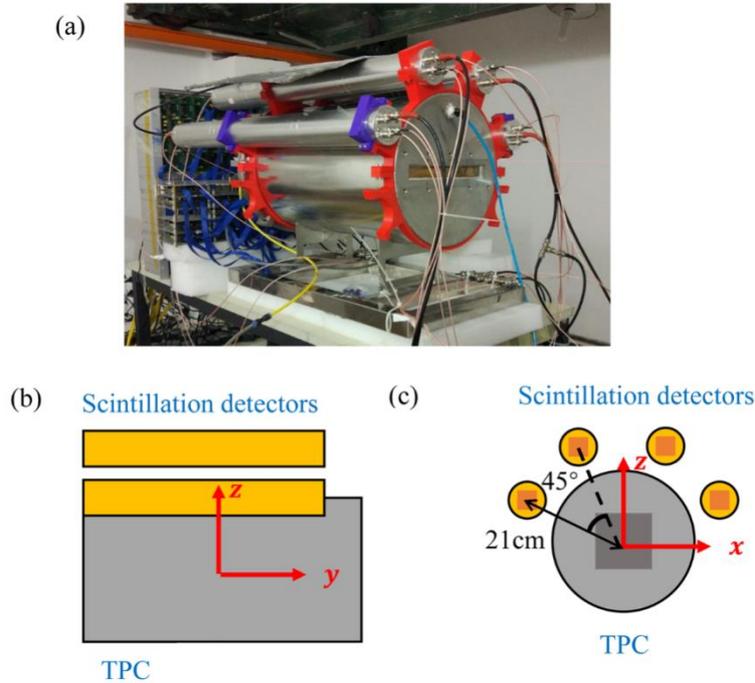

**Figure 3.** Design of the prototype system

### 2.1 The neutron TPC

The TPC is modified from our former nTPC [12], which is a fast neutron spectrometer based on the elastic scattering of $^1$H nucleus in the working gas. The outer size of the TPC is $\Phi 35\times 55$ cm$^3$, and the sensitive volume is $10\times 10\times 50$ cm$^3$. The field cage is a flexible PCB fixed on the outer face of a $\Phi 30\times 50$ cm$^3$ PMMA cylinder. Ar-$C_2H_6$ (50-50) is chosen as the working gas for a high proportion of hydrogen and good electron drift characteristics. The operating pressure is 1 atm. After being multiplied by a $10\times 10$ cm$^2$ triple-GEM detector, the signals are collected on readout pads connected to 36 different 16-channel ASIC boards (576 channels in total). Each channel is sampled by a 25 MHz FADC.

The simulation results show that the angular resolution of the system worsens with a pad size of more than 4×4 mm$^2$ [9]. To reduce the number of electronic channels, 4×4 mm$^2$ is chosen as the readout pad size.

The field cage and stainless vessel may interact with the scattered neutrons before they reach the scintillation detectors. According to the design of nTPC, the thicknesses of the stainless vessel and PMMA cylinder in the field cage are 2 and 5 mm, respectively. The transmission ratio of the scattered neutrons with different energy is calculated using the neutron cross-sections (from ENDF[13]) as shown in Figure 4, which indicates the ratio of scattered neurons passing through without being scattered by the stainless vessel and PMMA cylinder. The result shows that the transmission ratio is more than 70 % for scattered neutrons with energy more than 0.5 MeV.

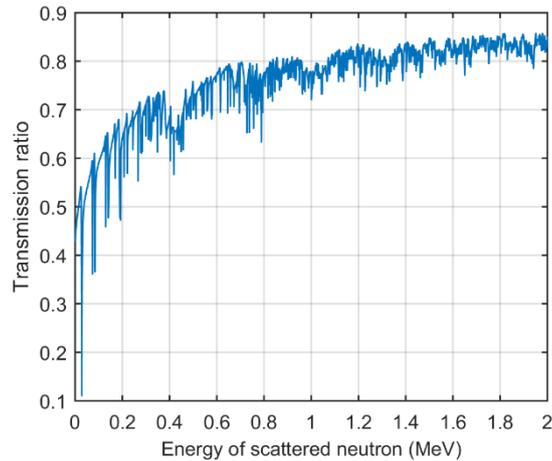

**Figure 4.** Transmission ratio of the scattered neutron (the valleys result from the neutron resonance peaks)

### 2.1.1 Correction of the electron drift velocity

The electron drift velocity is of great importance in track reconstruction of the TPC. Previous works [12] showed that the simulated value deviated from the experimental value. The correction is based on the fact that the flight time of the scattered neutron (~ several tens of nanoseconds) is negligible compared to the drift time in the TPC (~ several microseconds), so the difference between the trigger time of the two detectors in a coincident event must be between 0 and the maximum drift time of the TPC.

In Figure 5(a), the plateau results from the chance coincident events. The maximum drift time is determined by the two edges of the plateau. The chance coincident events are distributed evenly in the range of approximately 22 μs, which was the data acquisition time of the scintillation detectors. Each edge is fitted to a Gaussian function after being differentiated, as shown in Figures 5(b) and (c). Then, the maximum drift time is calculated as 14.3 ± 0.4 μs. The maximum drift length of the TPC is 50 ± 0.2 cm, the uncertainty of which is determined by the machining accuracy. Thus, the electron drift velocity is calculated as 3.49 ± 0.11 cm/μs. Moreover, the absolute energy deposition position in the drift direction can be calculated from the electron drift velocity and the trigger time difference between the TPC and the scintillator detector.

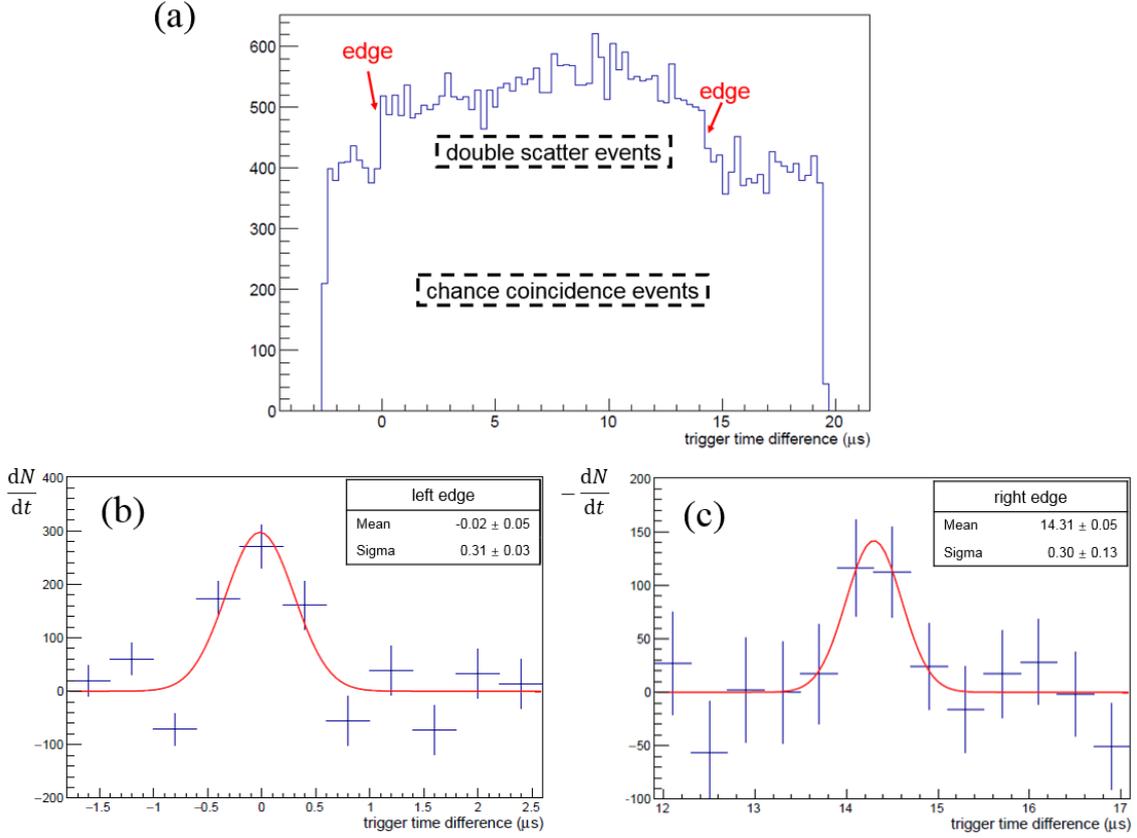

**Figure 5.** (a) Histogram of the trigger time difference. (b) The fitting of the left edge (y-axis: differentiation of the histogram). (c) The fitting of the right edge (y-axis: differentiation of the histogram).

### 2.1.2 Track reconstruction

The uncorrelated noise around the track are eliminated using the Hough transform, as shown in Figure 6. First, the points are projected from the x-z plane onto the ρ-θ plane [12].

$$\rho(\theta) = x \cos\theta + z \sin\theta, \left( -\frac{\pi}{2} < \theta < \frac{\pi}{2}, |\rho| \leq \sqrt{z_{\max}^2 + x_{\max}^2} \right) \quad (1)$$

Then, the intersection that is crossed by the most curves is selected. Finally, the points of the track are selected by retaining the curves passing through the intersection point.

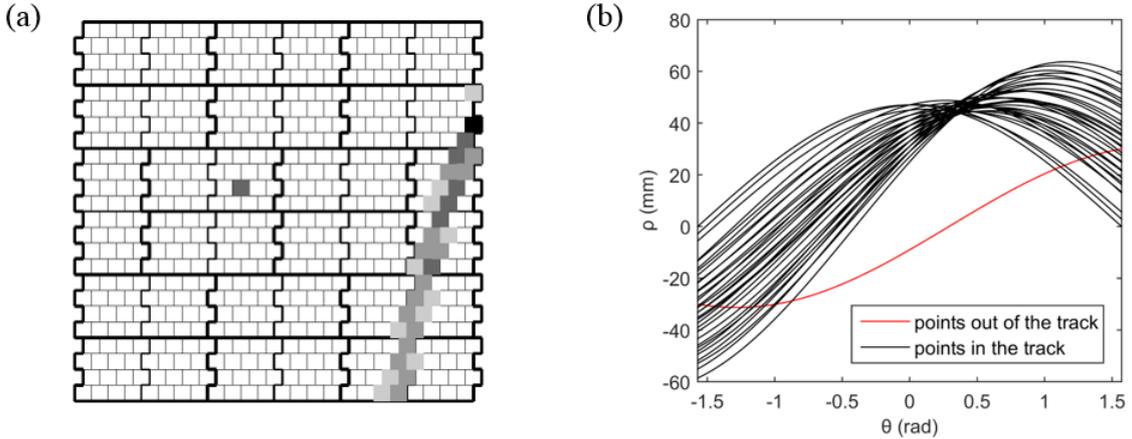

**Figure 6.** Hough transform
(a) Hit map of an event (b) The curves after Hough transform

The track direction is reconstructed by fitting the pad hits, as shown in Figure 7. A weighted least square fitting is used to fit the track points in the x-y and x-z (or y-z) planes. The distances between the track points and the fitting line are used as the residuals for the fitting in the x-y plane to suppress the reconstruction bias.

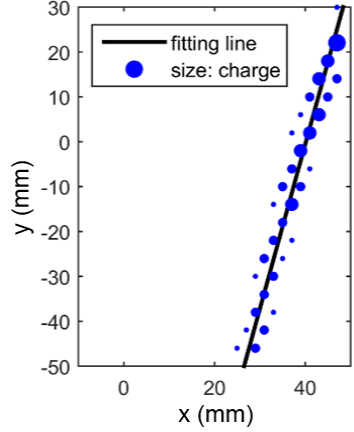

**Figure 7.** Schematic diagram of the weighted least square fitting

The track is determined by the center of gravity (COG) of the pad hits and the reconstructed direction. The total energy deposition is represented by the total charge collection.

The starting point of the track is used in the reconstruction of the direction of the scattered neutron, thus it must be reconstructed with high accuracy. In our previous works, the vertices of the track were roughly reconstructed as the two hit points with the longest distance between them.[12] However, the reconstruction accuracy was low owing to the bias resulting from the diffusion in the working gas.

A method based on the $dE/dx$ is shown in Figure 8. The method includes the following steps: (1) adding up the total charge collected by the pads in a pad row; (2) representing the $dE/dx$ value along a track according to the total charge of each triggered pad row (the blue line); (3) indicating the end of the track by the Bragg peak (the right end); (4) determining the starting position by an interpolation method at the start of the track (shown in red); (5) determining the end position by a similar process (shown in black); and (6) determining the track length by the starting and end positions.

This method is still somewhat rough (e.g., the maximum $dE/dx$ in the first half of the track is used as a rough estimate of $dE/dx$ of the starting point) and is not unbiased; however, the reconstruction bias is reduced compared with the previous method. The reconstruction accuracy is evaluated using the experimental data in Section 4.1.1.

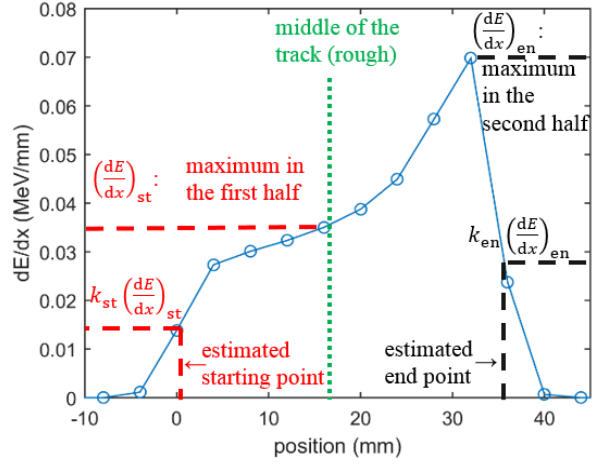

**Figure 8.** Reconstruction of the starting point and track length (simulation result, the interpolation parameters $k_{\text{st}}$ and $k_{\text{en}}$ are determined by Monte Carlo simulation using Geant4 and ROOT[9])

### 2.1.3 Calibration of the gain uniformity

The difference in gains among different pads mainly results from the non-uniformity of the triple-GEM module and the readout electronics.

The gain of the electronics readout channels was calibrated using a pulser. Except for five dead channels, the gain inconsistency of the other 571 readout channels was estimated as 7.7% (relative standard deviation).

The non-uniformity of the GEM gain was calibrated by a cosmic muon test. The $dE/dx$ value of the cosmic muon follows the Landau distribution (Figure 9(a)). The most probable values of the Landau distribution measured at different pads are used to calibrate the GEM gain. The test result shows that the gain inconsistency caused by gain fluctuations of the triple-GEM module is 10.6% (relative standard deviation), as shown in Figure 9(b).

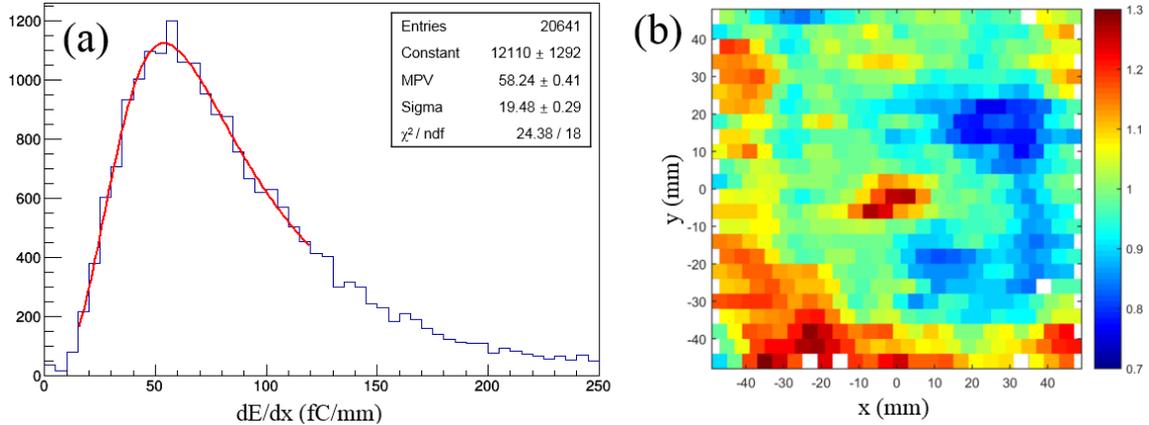

**Figure 9.** Calibration of the GEM gain uniformity. (a) The histogram of the $dE/dx$ value at one of the pads. (b) The calibration result of the relative GEM gains.

### 2.2 The scintillation detectors

As shown in Figure 10, a dual-end readout is used to measure the depth of interaction (DOI), which is the lengthwise position of the energy deposition in the scintillator bar. Each scintillation detector is coupled with two Hamamatsu CR105-01 PMTs at both ends. Four 1.6×1.6×40 cm³ SP101 plastic scintillator bars are used in a 2×2 array to achieve a better DOI resolution. Tyvek paper is used as a diffuse reflector. The PMTs are connected to homemade pulse shaping circuits and subsequently sampled by a CAEN V1724 ADC. The DOI information is calculated from the signal amplitudes by Equation (2) [10][14]:

$$\text{DOI} = \frac{1}{2\alpha} \ln \frac{E_2}{E_1} \qquad (2)$$

where $\alpha$ is the light attenuation coefficient, and $E_1$ and $E_2$ are the signal amplitudes. The optimal value of $\alpha$ is $2.9/L$ for the best DOI resolution, where $L$ is the length of the scintillator[14]. The α value of the $2 \times 2$ array scintillator is closer to the optimal value (i.e., better DOI resolution) than that of a single wider scintillator setup.

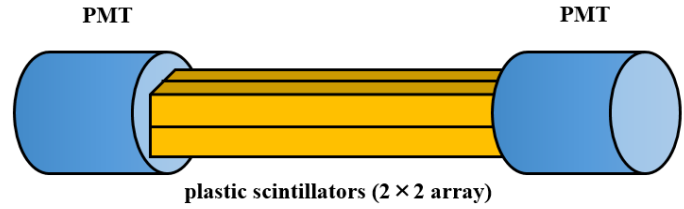

**Figure 10.** Design of the dual-end readout scintillation detector

The scintillation detectors were tested using collimated [137]Cs and [60]Co sources at seven different positions to estimate the α value. The α value is fitted to be 0.00368 mm[-1] (Figure 11(a)). The DOI resolution ($\sigma_{\text{DOI}}$) is calculated according to the resolution of $E_2/E_1$ and Equation (2), and is estimated to be 37 mm at 100 keVee (keV equivalent electron), as shown in Figure 11(b). The reconstruction bias of the DOI is estimated to be less than 4 mm among all collimated positions by comparing the experimental data with the fitting line in Figure 11(a).

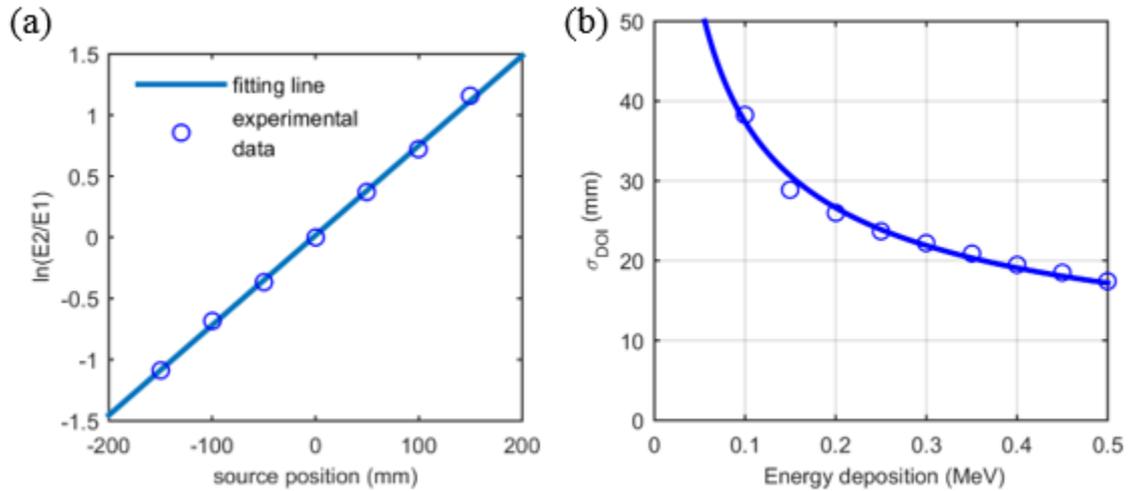

**Figure 11.** Test result of the scintillation detector. (a) The reconstruction of the DOI acoording to Equation (2). (b) The DOI resolution of the scintillation detectors.

## 3. Estimation of the performance using double scatter events

### 3.1 The angular resolution measure

Similar to Compton imaging [15], the resolution of the angle between the projection plane $\lambda$ and the actual direction of the incident neutron (the resolution of angle $\delta$ in Figure 12) is defined as the angular resolution measure (ARM).

– 8 –

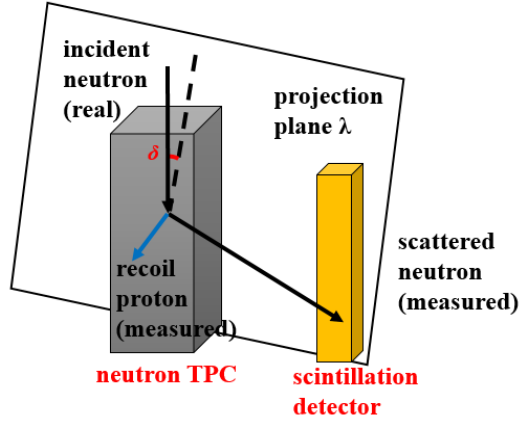

**Figure 12.** Defination of the angular resolution measure.

### 3.1.1 Analysis of the key factors affecting angular resolution measure

Because angle $\omega$ is 90 °, the ARM is

$$\sigma_\delta = \sqrt{\sigma_{\delta_p}^2 \cos^2\alpha + \sigma_{\delta_n}^2 \sin^2\alpha} \tag{3}$$

where $\alpha$ is the recoil angle, $\sigma_{\delta_p}$ is the resolution of the angle ($\delta_p$) between the true direction of the recoil proton and the projection plane $\lambda$, and $\sigma_{\delta_n}$ is the resolution of the angle ($\delta_n$) between the true direction of the scattered neutron and the projection plane $\lambda$.

The track of the recoil proton is reconstructed by the TPC. Two factors contribute to $\delta_p$: the multiple Coulomb scattering of the protons ($\sigma_{\text{scattering}}$), and the fitting accuracy of the track ($\sigma_{\text{fit}}$). The resolution of $\delta_p$ is given by

$$\sigma_{\delta_p} = \sqrt{\sigma_{\text{scattering}}^2 + \sigma_{\text{fit}}^2} \tag{4}$$

$\sigma_{\text{scattering}}$ is related to the composition of the working gas and track length. $\sigma_{\text{fit}}$ is related to the track length and readout pad size, and is also affected by the diffusion coefficient of the working gas.

$\sigma_{\delta_n}$ is determined by the travel distance of the scattered neutron and position resolution of the detectors. The position resolution of the TPC is negligible compared to that of the scintillation detector. The DOI resolution of the scintillation detector is much worse than the position resolution in the other two directions. To improve the angular resolution, an event reconstruction method based on the fact that $\omega$ equals 90 °is developed to calculate the projection plane without the DOI information. In this case, the DOI information is only used to calculate $\omega$ and distinguish the chance coincidence background.

The ARM is discussed in detail in Appendix A, where the analytical expressions of $\sigma_{\text{scattering}}$, $\sigma_{\text{fit}}$, and $\sigma_{\delta_n}$ are deduced for further estimation.

### 3.1.2 Calculation results

Based on the analysis and equations in Appendix A, the ARM is calculated using the parameters of the prototype system, which shows the limit of the angular resolution of the prototype system.

### 3.1.2.1 $\sigma_{\delta_p}$

$\overline{\sigma_{\delta_p}}$ can be calculated using Equations (A2), (A6), and (A15). Some parameters used in the calculation are from [16] and listed in Table 1.

**Table 1** Some parameters used in the calculation of $\overline{\sigma_{\delta_p}}$

| Parameter | Description | Value | Comment |
|---|---|---|---|
| $\overline{W}$ | Average ionization energy of the working gas | 26 eV | |
| $\alpha_{\text{Polya}}$ | Polya distribution factor of the GEM module gain | 0.5 | |
| $D_L$ | Longitudinal diffusion coefficient | 231 μm/cm$^{1/2}$ | Garfield simulation result |
| $D_T$ | Transverse diffusion coefficient | 293 μm/cm$^{1/2}$ | Garfield simulation result |
| $\sigma_{z_0}$ | Intrinsic z resolution of the detector and electronics system | 300 μm | Experimental result of a cosmic muon test, as the residual in a linear fit of the muon track (relative resolution). |

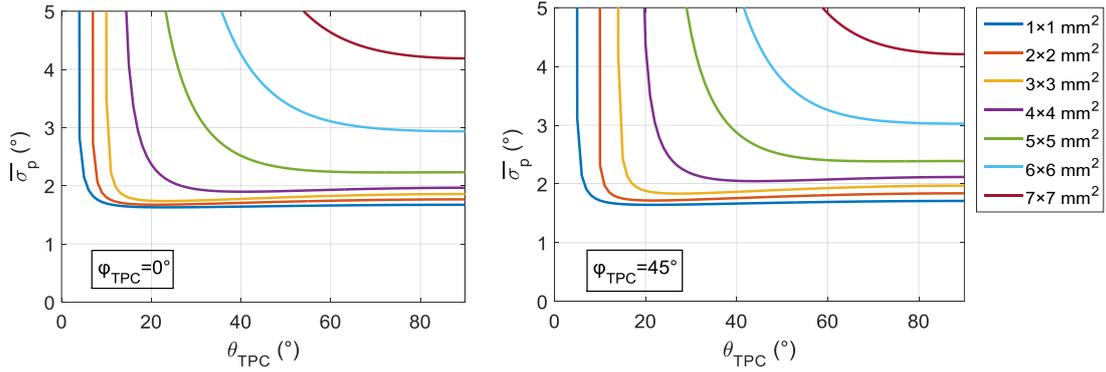

**Figure 13.** Calculation result of $\overline{\sigma_{\delta_p}}$ for different pad sizes. $\theta_{\text{TPC}}$ and $\varphi_{\text{TPC}}$ are the zenith and azimuth angles of the recoil proton direction in the TPC coordinate system (see Appendix A.2).

The pad size is a crucial design parameter related to both the angular resolution and the detection efficiency. For protons with an energy of 1 MeV, which is the average energy for recoil protons produced by fission neutrons, the estimated $\overline{\sigma_{\delta_p}}$ of different pad sizes are shown in Figure 13.

When the pad size is less than or equal to 4×4 mm$^2$, $\overline{\sigma_{\delta_p}}$ is around 2° and mainly results from the multiple Coulomb scattering when $\theta_{\text{TPC}}$ is above 30°. This result is in good agreement with the Monte Carlo simulation results [10] that the angular resolution of the system has no significant difference with readout pad sizes smaller than 4×4 mm$^2$.

### 3.1.2.2 $\sigma_{\delta_n}$

$\sigma_{\delta_n}$ is calculated from the position resolution of the scintillation detectors (refer Appendix A.2). DOI resolution used for the calculation is 37 mm (test result for 100 keVee). Even with some improvement in the system design, the ARM calculated using the DOI information is not good in

– 10 –

a wide range of projection plane directions (Figure 14 (a), only results with 0<$\theta_{proj}$<90° and 0<$\varphi_{proj}$<90° are shown due to the symmetry (refer Appendix A.3)).

The DOI resolution can be further improved by reducing the length of the scintillator, but this results in a worse efficiency. By reconstructing the plane $\lambda$ using the angle $\omega$ instead of the DOI information, a good resolution can be achieved without sacrificing efficiency (Figure 14 (b)).

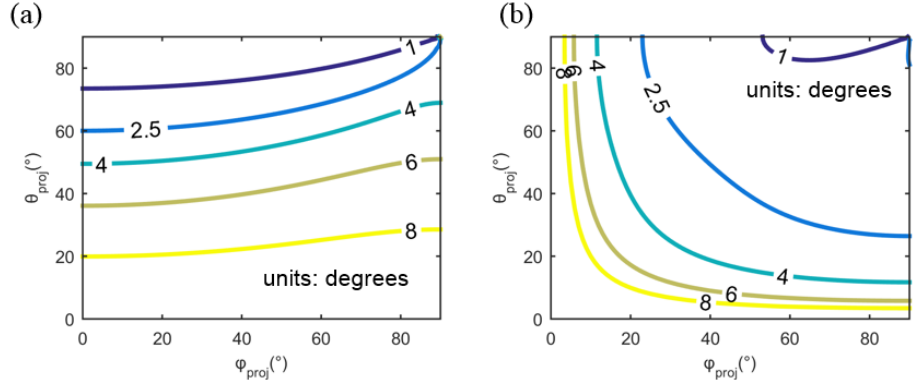

**Figure 14.** Contour plot of the estimated $\sigma_{\delta_n}$ vs. direction of the projection plane (a) Calculated using the DOI information (Equation (A24)); (b) Calculated using $\omega = 90°$ (Equation (A27), $\sigma_\alpha = 2°$ according to Section 3.1.2.1).

### 3.1.2.3 The angular resolution measure

The ARM is calculated from the estimated $\sigma_{\delta_p}$ and $\sigma_{\delta_n}$. According to Section 3.1.2.1, $\overline{\sigma_{\delta_p}}$ of 2° is used for the calculations. As shown in Figure 15, the ARM of the prototype system is better than 2.5° in the range 30°<$\theta$ < 150°, 40°<$\varphi$ < 140°. Assuming that the projection plane is evenly distributed in all directions, 77% of the neutron events can be reconstructed with an ARM better than 2.5°. The ARM of the neutron imaging experiment in Section 4 ($\theta = 90°$, $\varphi = 90°$) is estimated to be 1.9°.

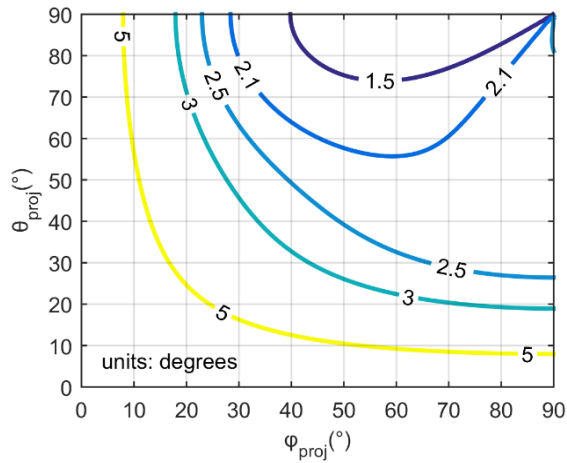

**Figure 15.** Contour plot of the estimated angular resolution measure (calculated using $\omega = 90°$)

## 3.2 The imaging efficiency

The neutron TPC has high efficiency and a 4π FOV using single scatter events. Adding plastic detectors breaks the symmetry and reduces the efficiency. The efficiency varies with the incident neutron direction. It is affected by the energy threshold of the TPC and scintillation detector. The efficiency are estimated by numerical calculations according to the system design and neutron cross-sections from ENDF [13]. The efficiency for fission neutrons from different directions is shown in Figure 16. The low efficiency of neutrons in the upper side results from the location of the scintillation detectors, which is limited by the design of the nTPC.

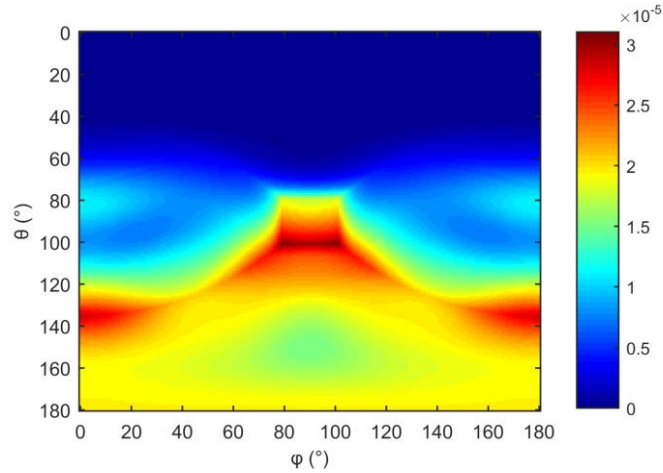

**Figure 16.** The imaging efficiency for fission neutrons from different directions (the low energy thresholds for the TPC and scintillation detector are 1 and 0.4 MeV, respectively)

The efficiency for different energy thresholds is shown in Figure 17. The efficiency can be further improved by using more scintillation detectors.

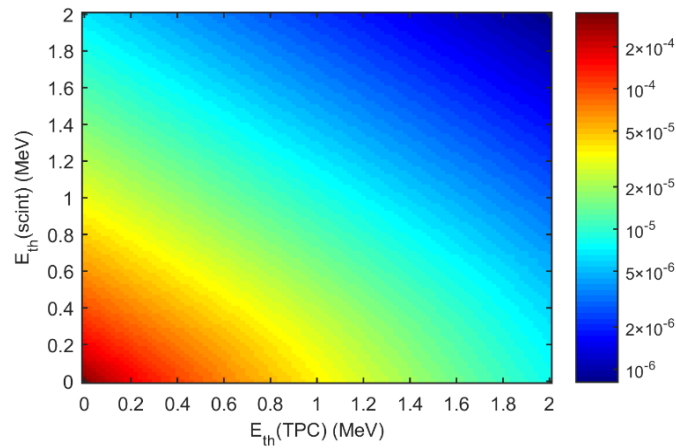

**Figure 17.** The imaging efficiency for different energy thresholds ($\theta = 90°$, $\varphi = 90°$)

## 4. The neutron test and the imaging reconstruction

### 4.1 Experimental setup and data analysis

The neutron imaging experiment was carried out in the Laboratory of Metrology and Calibration Technology, China Institute of Atomic Energy. A $^{252}$Cf source with a neutron production rate of $5.96 \times 10^6 \, \text{s}^{-1}$ was placed approximately 2 m away from the prototype system close to the axis of the TPC, as shown in Figure 18. The precise measurement of the position was not carried out owing to the high radioactivity of the source. The design of the prototype system is shown in Table 2. The prototype system roughly imaged the neutron source using the TPC only to determine the region of interest and then reconstructed a high-resolution image using the double scatter events.

**Table 2** Key parameters in the neutron imaging experiment

| | Parameter | Design |
|---|---|---|
| TPC | Working gas | Ar-C$_2$H$_6$ (50-50) 1 atm |
| | Sensitive volume | $10 \times 10 \times 50$ cm$^3$ |
| | Drift field | 174 V/cm |
| | Readout detector | Triple-GEM |
| | Readout pad size | $4 \times 4$ mm$^2$ |
| Plastic scintillation detector | Scintillator type | SP101 |
| | Scintillator size | $1.6 \times 1.6 \times 40$ cm$^3$ per element, $2 \times 2$ elements array for a single detector |
| | Reflector | Tyvek paper, 0.2 mm thick |
| | PMT type | Hamamatsu CR105 |
| | PMT size | $\Phi$ 2 inch, 2 PMTs for a single detector (dual end) |

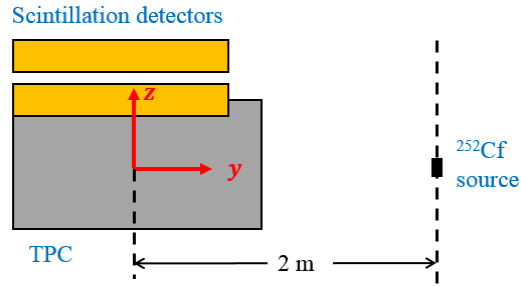

**Figure 18.** Setup of the neutron imaging test

### 4.1.1 Selection of proton tracks in the TPC

The experimental data of the TPC contains a large number of particle tracks. Besides the recoil protons, there are two main kinds of background particles: electrons and $^{12}$C/$^{40}$Ar nuclei. The electrons are mainly produced by the nuclear reaction associated gamma rays. Due to their much smaller $dE/dx$ value, the electrons have longer track lengths and smaller energy deposits than the protons. The $^{12}$C/$^{40}$Ar nuclei are recoiled by neutrons with much lower energy due to their larger masses compared to the protons. Considering their larger $dE/dx$ value, the track lengths of the $^{12}$C/$^{40}$Ar nuclei are much shorter than protons. Therefore, both electrons and $^{12}$C/$^{40}$Ar nuclei can be distinguished by deposition energy and track lengths (Figure 19).

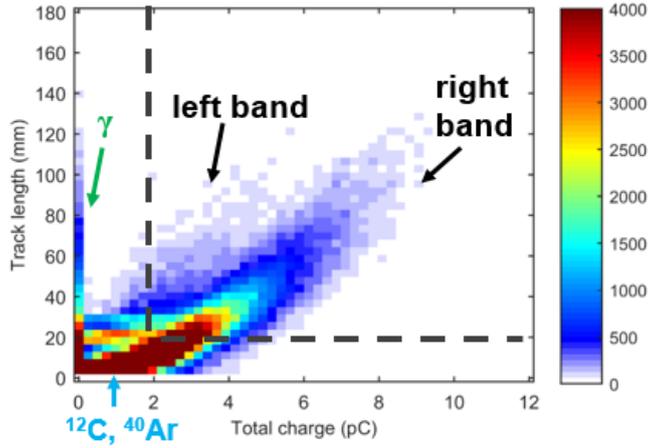

**Figure 19.** 2D histogram of the particles recorded in the experiment

The absolute energy scale is estimated by comparing the calculated track length and the Monte Carlo simulation result by the following steps: (1) determining the relationship between the track length and energy (i.e. $L_{\text{simu}}(E)$) by a Geant4 simulation (Figure 20(a)); (2) estimating the value and resolution of the track length by choosing a narrow charge window and fitting the histogram of the track length values within the window by Gaussian function (the real resolution is smaller because the charge resolution will contribute to this estimation); and (3) fitting the experimental data ($L_{\text{exp}}(Q)$) using the function (Figure 20(b)).

$$L_{\text{exp}}(Q) = L_{\text{simu}}(k_{\text{scale}}Q) + L_{\text{bias}} \quad (5)$$

The absolute energy scale ($k_{\text{scale}}$) and the reconstruction bias of the track length ($L_{\text{bias}}$) are estimated as 0.452 MeV/pC and 1.8 mm, respectively.

The fitted standard deviation of the track length is relatively small at low energies (1.7 mm at 1.8 pC), but increases with higher energy of the recoil proton owing to the growth of contributions from the zenith angle resolution and energy resolution. The derived track length is longer than the real value (positive bias) because of the diffusion of the charge in the gas. The bias of the previous method is usually ~6 mm. It is improved using the new method (Figure 8). Assuming the bias and uncertainty of the starting point equal to those of the end point, the bias and uncertainty are estimated to be less than 0.9 and 1.2 mm (half of the bias and $\sqrt{0.5}$ of the fitted standard deviation at 1.8 pC), respectively, which is much lower than those of the interaction point in the scintillation detector estimated in Section 2.2.

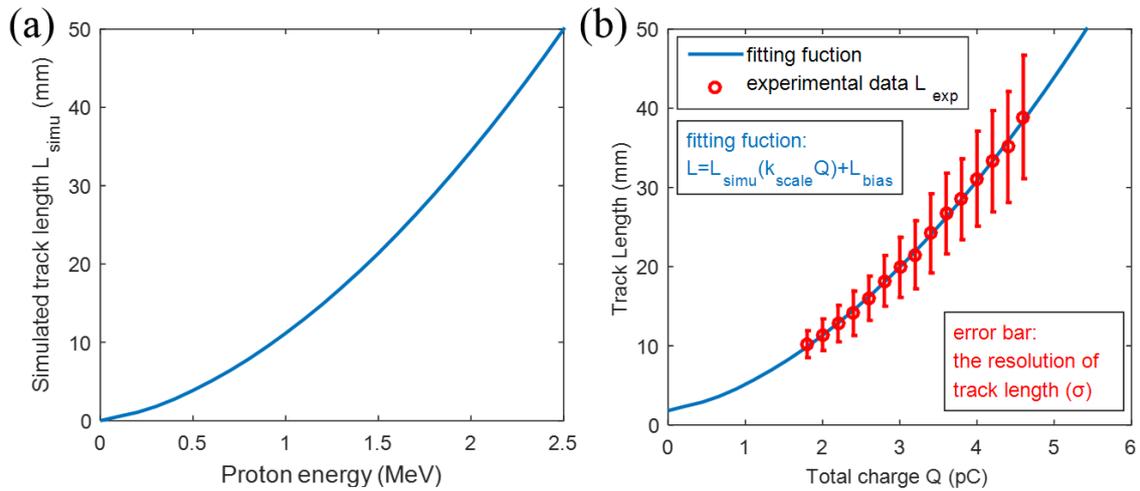

**Figure 20.** Estimation results using the track length distribution at different total charge. (a) Geant4 simulation result of the track length. (b) Estimation of the absolute energy scale and the reconstruction bias of the track length.

The threshold is set to 18 mm for the track length and 2 pC (0.9 MeV) for the total charge to achieve a good angular resolution, which results in a selection efficiency of 0.38 for the recoil protons. The $^{12}$C/$^{40}$Ar events are negligible after selection according to Monte Carlo simulation by Geant4: the minimum energy for a recoil $^{12}$C nuclei to exceed the track length threshold is 18 MeV, which means a selection efficiency of less than $10^{-18}$. The selected proton tracks have a double band structure; the track of a proton in the right band is entirely in the sensitive volume, while the track of a proton in the left band starts in the sensitive volume but ends outside. The proton tracks from both bands are selected for further analysis because their directions and starting positions can be reconstructed for fast neutron imaging.

### 4.1.2 Selection of the double scatter events

One of the most widely used techniques in coincidence measurements is the time window. As shown in Figure 5, a time window with the width of the maximum drift time in the TPC can be used to distinguish the chance coincident events, but it is not strict enough because the drift time of the TPC is too long. The value of angle $\omega$ (90° for double scatter events) can be used to distinguish the chance coincident events. [11] As shown in Figure 21, by using an angle window for $\omega$ from 85 to 95°, the double scatter events are distinguished effectively from the chance coincident events. The true to coincidence ratio of the double scatter events increases from 0.056 to 0.236 after using the angle window, which is calculated from the net and total counts of the experimental results. The selection efficiency of the angle window is related to the DOI resolution and is calculated to be 0.68 and 0.84 for energy depositions of 100 and 200 keVee, respectively (estimated using the angle window and test result of the DOI resolution in Figure 11).

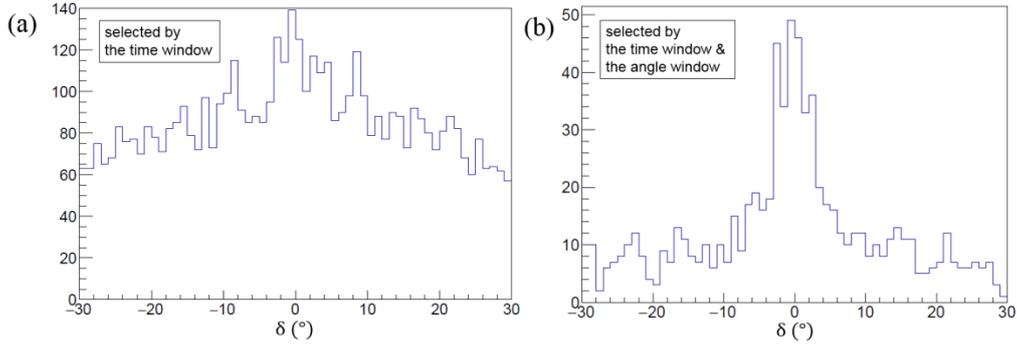

**Figure 21.** Histograms of the angular resolution measure.
(a) Selected by the time window. (b) Selected by both the time window and the angle window.

### 4.2 Imaging reconstruction

The first imaging results were presented in a previous report[9]. The prototype system achieved an angular resolution of 91 °(FWHM) and an efficiency of $7.1 \times 10^{-3}$ using single scatter events in the TPC, which allowed for fast imaging of the hot spot with a $4\pi$ FOV.

The more accurate imaging results are achieved using the double scatter events. In this section, three image reconstruction algorithms widely used in Compton imaging [17] and medical imaging [18] are applied to the neutron imaging system to improve the angular resolution. If the direction of the incident neutron can be determined as where the measurement error reaches the minimum in the projection plane $\lambda$, the ARM equals to the angular resolution. The angular resolution is worse than the ARM because the direction of the incident neutron are limited to one quarter of the projection plane $\lambda$. The angular resolution of the reconstructed image varies with the image reconstruction algorithms.

### 4.2.1 Simple back-projection

The simple back-projection (SBP) method is one of the most straightforward algorithms. The direction of the incident neutron is limited to part of a plane (Figure 2), and the probability can be calculated for each possible direction. The SBP image is calculated by simply summing up the probabilities of all the measured events. The SBP method can be performed event by event, but the image is blurred because the back-projection planes overlap with each other.

Figure 22 is an image reconstructed using the SBP method. The streak artifacts are caused by the asymmetry of the positions of four scintillation detectors. The angular resolution of the image is 7.8 °(FWHM). The efficiency is calculated to be $2.2 \times 10^{-5}$ [9], which is in good agreement with the results calculated from the cross-sections (efficiency is between $2.93 \times 10^{-5}$ and $1.94 \times 10^{-5}$ for $E_{scin}(th)$ between 100 and 500 keV when $E_{TPC}(th)$ is 1 MeV (refer Figure 17)).

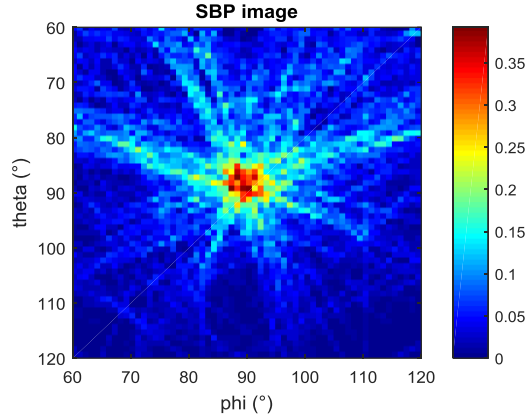

**Figure 22.** SBP image of the double scatter events.

For the elastic scattering between a fast neutron and a proton, the distribution of the recoil angle $\alpha$ is proportional to $\sin 2\alpha$. Then the point spread function (PSF) of the SBP image can be calculated as:

$$h_{\text{PSF}}(\cos\psi_0) = \frac{1}{2\pi\ \sin\psi_0}\int_0^{\frac{\pi}{2}-\alpha} 2\sin 2\alpha \sin 2\left(\frac{\pi}{2}-\psi_0-\alpha\right)d\alpha$$
$$= \frac{(\pi-2\psi_0)\cos 2\psi_0 + \sin 2\psi_0}{4\pi\sin\psi_0} \quad (6)$$

where $\psi_0$ is the angle between the source direction and the direction of the image pixel. The imaging result using the SBP method can be deduced using the ARM:

$$h_{\text{SBP}}(\Omega) = \int Gauss(0,\sigma_\delta,\psi) h_{\text{PSF}}(\cos\psi) d\Omega' \quad (7)$$

where $\psi$ is the angle between $\Omega$ and $\Omega'$, and $Gauss(0,\sigma_\delta,\psi)$ is the Gaussian distribution of $\delta$. Equation (7) has no analytical solution due to the integral of the Gaussian function, but the angular resolution of the SBP image can be determined from the ARM as the following equation using numeral calculations:

$$\text{FWHM}_{\text{SBP}} \cong 3.74\ \sigma_\delta \quad (8)$$

The angular resolution of the SBP image is estimated to be 7.1 ° using Equation (8) and the estimation result of the ARM in Section 3.1, which is in good agreement with the experimental result.

### 4.2.2 Filtered back-projection

The filtered back-projection (FBP) method is an algorithm proposed to improve the performance of the SBP method. The blurring is filtered out in spherical harmonics domain due to the point spread function of the SBP image. The FBP image $g(\Omega)$ can be calculated from the SBP image $g'(\Omega)$ using the following equation [17]:

$$g(\Omega) = \int g'(\Omega')h^{-1}(\cos\psi) d\Omega' \quad (9)$$

where $h^{-1}(\cos\psi)$ is defined as:

$$h^{-1}(\cos\psi) = \sum_{n=0}^{\infty}\left(\frac{2n+1}{4\pi}\right)^2 \frac{1}{H_n} P_n(\cos\psi) \quad (10)$$

where $\psi$ is the angle between $\Omega$ and $\Omega'$, and $H_n$ are the coefficients of $h_{\text{PSF}}(\cos\psi_0)$ expanded on the Legendre polynomials.

The function $h^{-1}(\cos\psi)$ can be approximated using Equation (10) by adding up the first $n^{th}$ elements of the infinite series. With the pre-calculation of $h^{-1}(\cos\psi)$, the FBP method can be performed in the event by event mode:

$$g(\Omega)_N = \int g'(\Omega') \left[\sum_{n=0}^{N} \left(\frac{2n+1}{4\pi}\right)^2 \frac{1}{H_n} P_n(\cos\psi)\right] d\Omega' \quad (11)$$

When $N$ approaches infinity, the angular resolution of the FBP image approaches its limit, i.e., the ARM or 2.355 times the ARM in the form of FWHM (calculated by Equation (9)). However, the cutoff of the infinite series results in the deterioration of the angular resolution, which is more serious with a larger $N$. This deterioration may result from factors such as the asymmetry of the system and limited counts of neutron events. Therefore, the value of $N$ used for the reconstruction should be carefully adjusted, as shown in Figure 23.

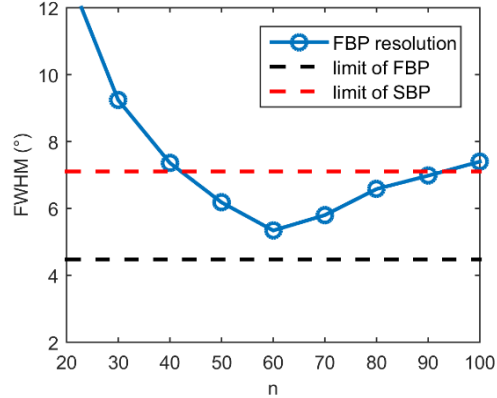

**Figure 23.** The resolution of FBP images reconstructed with different values of $N$

The FBP image of the experimental data is shown in Figure 24. The artifact is smaller than the SBP image, but does not disappear due to the asymmetry in the prototype system. The angular resolution is improved to 5.3 °(FWHM) using the same data from Figure 22.

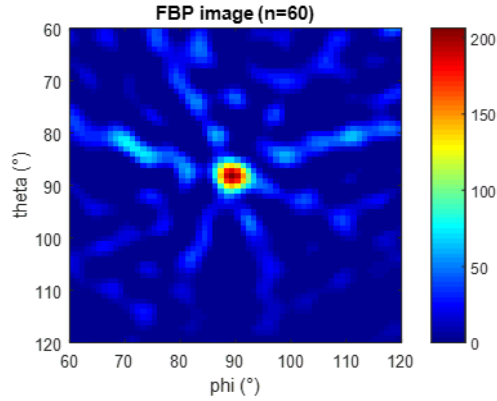

**Figure 24.** FBP image of the double scatter events ($N = 60$).

### 4.2.3 Maximum likelihood expectation maximization

The maximum likelihood expectation maximization (MLEM) method is an iterative algorithm to calculate the source distribution with the maximum likelihood of the measured data. The iteration is performed by the equation: [17]

$$\lambda_j^{n+1} = \frac{\lambda_j^n}{s_j} \sum_i \frac{Y_i t_{ij}}{\sum_k t_{ik} \lambda_k^n} \qquad (12)$$

where $\lambda_j^n$ is the value of direction $j$ at the $n^{\text{th}}$ iteration, $s_j$ is the detection efficiency of neutrons from direction $j$, $Y_i$ is the number of measurement $i$, and $t_{ij}$ is the probability of a neutron from direction $j$ to be measured as measurement $i$. $t_{ij}$ is the most essential value of all the parameters and is calculated by:

$$t = t_{\text{energy}} \cdot t_{\text{scatter}} \cdot t_{\text{direction}} \cdot t_{\text{efficiency}} \qquad (13)$$

where $t_{\text{energy}}$ is the probability of the source to emit a neutron with the estimated energy, $t_{\text{scatter}}$ is the probability for the neutron to be scattered by $^1$H in the TPC, $t_{\text{direction}}$ is the probability for the scattered neutron to be in the direction of the second scattering position, and $t_{\text{efficiency}}$ is the probability for the scattered neutron to be detected by the scintillation detector.

The MLEM method can reconstruct the image precisely after a number of iterations, but there are also some disadvantages. The MLEM method cannot reconstruct the image in the event by event mode, and the iterations result in a high computational cost.

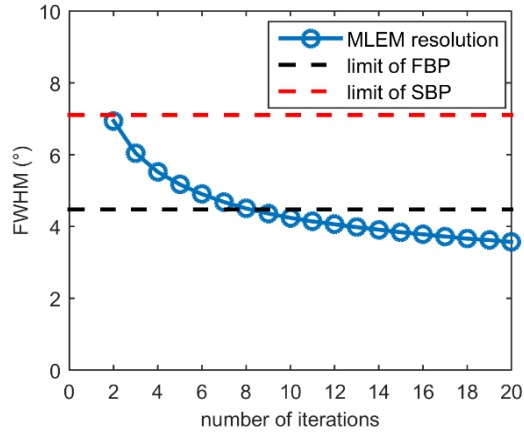

**Figure 25.** The resolution of the MLEM image after different number of iterations

The resolution increases with every iteration, but more computational time is required. After several iterations, the resolution of the MLEM image is better than the FWHM limits of the SBP and FBP algorithms, as shown in Figure 25.

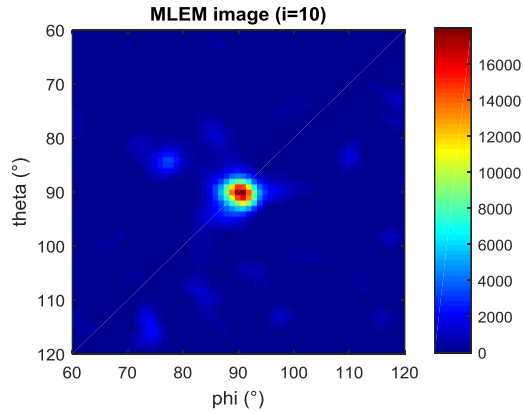

**Figure 26.** MLEM image of the double scatter events after 10 iterations.

Figure 26 is the MLEM image after 10 iterations. The angular resolution is 4.2 °(FWHM) and can be further improved with more iterations. There is no significant artifact compared to the FBP and SBP images.

### 4.3 Estimation of the imaging reconstruction bias of the prototype system

The imaging reconstruction bias is of great importance since the goal of this paper is to develop a fast neutron imaging system with high-precision pointing accuracy. As shown in Figure 12, the projection plane is reconstructed by the directions of the recoil proton and the scattered neutron, and can be calculated by:

$$\Delta \delta = \Delta \delta_p \cos \alpha \pm \Delta \delta_n \sin \alpha \tag{14}$$

where $\pm$ depends on the direction of the two biases. The systematic errors of the two directions result in bias in the reconstruction image, which affects the pointing accuracy of the system.

#### 4.3.1 The systematic error of the recoil proton direction

Two factors contribute to the systematic error of the recoil proton direction: the drift velocity and fitting bias. The TPC coordinate system is used in the analysis in this section, with the drift direction of the TPC as the $z$-direction

The uncertainty of the drift velocity ($\sigma_{v_{\text{drift}}}$) results in a systematic error in the zenith angle of the reconstructed direction, which is estimated according to the error propagation:

$$\Delta \theta_{\text{TPC}}(v_{\text{drift}}) = \sin \theta_{\text{TPC}} \cos \theta_{\text{TPC}} \frac{\sigma_{v_{\text{drift}}}}{v_{\text{drift}}} \tag{15}$$

$\Delta \theta_{\text{TPC}}(v_{\text{drift}})$ reaches the maximum when $\theta_{\text{TPC}}$ is equal to 45 °, and is calculated as 0.87 ° using the estimation result of the drift velocity in Section 2.1.1.

The bias resulting from the fitting is estimated using Monte Carlo simulation. First, the energy deposition in the detectors is simulated by Geant4. Then, the drift and diffusion in the TPC is carried out by a ROOT program using a fast Monte Carlo method [19]. 1 MeV protons are randomly generated in the sensitive volume with random directions in the simulation. The gain uniformity of the pads (from Figure 9) is also included in the simulation. As shown in Figure 27, the fitting biases of the zenith and azimuth angles (the mean values in the figures) are simulated as 0.39 °and 0.16 °, respectively.

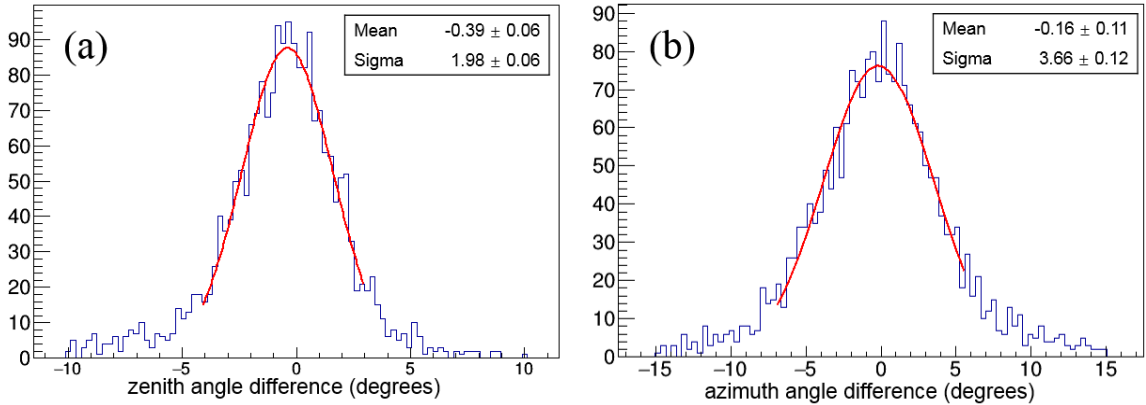

**Figure 27.** Histograms of the differences between the reconstruction directions and the true directions (simulation results). The sigma value represents the angular resolution of the track reconstruction, while the mean value represents the reconstruction bias.

### 4.3.2 The systematic error of the scattered neutron direction

The analysis in this section is based on the event reconstruction method using angle $\omega$ instead of the DOI information. According to the discussions in Appendix A.3 and Equation (A27), three factors contribute to the systematic error of the scattered neutron direction: the recoil angle in the TPC ($\alpha$), and the position of the two interaction points. Obviously, only the position errors in the direction perpendicular to the projection plane ($\Delta d_{\text{TPC}}$ and $\Delta d_{\text{scin}}$) can reduce the accuracy of the scattered neutron direction.

The systematic error of $\Delta \delta_n$ resulting from the position errors can be calculated as

$$\Delta \delta_n(d) = \frac{1}{L_{\text{travel}}} (\Delta d_{\text{TPC}} \pm \Delta d_{\text{scin}}) \tag{16}$$

where $L_{\text{travel}}$ is the travel length of the scattered neutrons between the two detectors.

The systematic error of $d_{\text{scin}}$ is 1 mm, which mainly results from the absolute position registration bias between the two detectors.

The systematic error of $d_{\text{TPC}}$ is deducted from the systematic error of the recoil proton track:

$$\Delta d_{\text{TPC}} = L_{\text{start\_to\_COG}} \tan \Delta \delta_p + \Delta l_{\text{COG}} \tag{17}$$

where $L_{\text{start\_to\_COG}}$ is the length between the starting point and COG of the track, and $\Delta l_{\text{COG}}$ and $\Delta \delta_p$ are the systematic errors of the COG position and track direction, respectively, in the direction perpendicular to the projection plane. Assuming $\Delta l_{\text{COG}}$ is much smaller than the first term and COG is approximately at the middle point of the track, Equation (17) can be approximated as

$$\Delta d_{\text{TPC}} = \frac{1}{2} L_{\text{proton}} \tan \Delta \delta_p \tag{18}$$

where $L_{\text{proton}}$ is the length of the recoil proton track. For recoil protons with energy less than 3 MeV, $\Delta d_{\text{TPC}}$ is calculated as less than 1.1 mm.

The systematic error of $\Delta \delta_n$ resulting from the error of the recoil angle ($\alpha$) can be calculated using Equation (A26). It can be proved by coordinate transformation that $\Delta \delta_p$ and $\Delta \alpha$ follow the equation:

$$\Delta \delta_p^2 + \Delta \alpha^2 = \Delta \theta_{\text{TPC}}^2 + \Delta \varphi_{\text{TPC}}^2 \sin^2 \theta_{\text{TPC}} \tag{19}$$

### 4.3.3 The estimation results

According to the discussions in Sections 4.3.1 and 4.3.2, the reconstruction bias caused by different factors was estimated for the neutron test and listed in Table 3. A maximum bias was also calculated by assuming that the biases from different factors are all in the same direction.

The reconstruction bias varies with the incident neutron direction owing to the asymmetry of the system. The average reconstruction bias and the reconstruction bias under the worst condition are also estimated.

Table 3 Estimation result of the maximum imaging reconstruction bias (units: degrees)

|  |  | The neutron test | The worst condition | Average for different incident directions |
|---|---|---|---|---|
|  | Recoil proton | 0.08 | 0.89 | 0.69 |
| Scattered neutron | Position resolution | 0.24 | 0.31 | 0.29 |
|  | Recoil angle ($\alpha$) | 0.31 | 0.80 | 0.62 |
| **Maximum total bias according to Equation (14)** |  | **0.56** ($\alpha = 82°$) | **1.42** ($\alpha = 51°$) | 1.14 |

## 5. Discussion

The theoretical model of the coincidence imaging system can be used to optimize the detectors in future applications. Analytical results show that optimization of the readout pads of the TPC does not improve the angular resolution significantly due to the multiple Coulomb scattering. The angular resolution can be improved by using scintillators with smaller sections or longer distance from the TPC, but both of them will result in lower efficiency. Using $\omega = 90°$ other than the DOI information in the scintillation detector can achieve a good angular resolution without sacrificing efficiency.

The angular resolution of the double scatter events tested using the prototype system is better than that of a typical neutron scatter camera. The theoretical model provides the limit of the angular resolution of traditional back projection based online reconstruction methods, which agrees well with the experimental data. Although the statistical iterative method can breakthrough this limit and further improve the angular resolution, it is a time consuming method. In typical applications, online reconstruction using the event-by-event mode is preferred.

In our system, the TPC itself can give a quick result that shows the rough direction of the hot spot. According to the calculated efficiency values towards different directions, one can adjust the bearing of the coincidence imaging system so that it can efficiently provide a fine measurement towards the direction of interest using double scatter events.

The reconstruction bias is estimated to be less than $0.56°$ for the neutron test and less than $1.42°$ under the worst condition, indicating high-precision pointing accuracy. The bias mainly results from the error of the drift velocity, and can be reduced by high-precision measurement of its value.

The absolute z position information of the TPC in the double scatter events can be used to reduce the parallax error. It is not used in the image reconstruction of the neutron test because the distance between the neutron source and the detectors is needed, which cannot be precisely reconstructed because of the limited number of double scatter events. However, the parallax error is of more importance for applications in which the source is close to the detector.

The TPC in the prototype system is modified from a former TPC [12] and not optimized for fast neutron imaging. The efficiency can be improved by using a shorter distance between the two types of detectors, using a thinner pressure vessel and field cage, improving the DOI resolution, or using more scintillators.

The chance coincidence events increase the background and worsen the resolution. The true to chance ratio of the double scatter events is lower than that of a typical neutron scatter camera because the coincidence time of the former is much longer than that of the latter. The start time of the drift in the TPC can be determined by the cathode signal and used to shorten the coincidence time. However, this is very challenging because the signal is very small.

## 6. Conclusion

The neutron TPC can locate the hot spot in neutron scatter imaging with a high detection efficiency, $4\pi$ FOV, and good neutron/gamma discrimination ability. By adding the plastic scintillation detectors, the angular resolution can be significantly enhanced using double scatter events. The high-resolution imaging is very important for the detailed measurement of the distribution within a hot spot.

A theoretical model is proposed to evaluate the performances of a coincidence imaging system. The calculated results agree well with the experimental data. The model can be used to optimize the detectors.

Although the TPC prototype is not completely optimized for neutron scatter imaging, an angular resolution of around 5 °(FWHM) is achievable using an event-by-event reconstruction mode. The 4π FOV of the TPC and high angular resolution of the coincidence imaging system are very important and desirable for practical applications. We suggest that the combined TPC and scintillator system is suitable for SNM detection.

## Appendix A: Deduction of the analytical expressions of the angular resolution measure

As shown in Section 3.1, the ARM is calculated from the resolution of the angle between the true direction of the recoil proton and the projection plane ($\delta_\text{p}$), resolution of the angle between the true direction of the scattered neutron and the projection plane ($\delta_\text{n}$), and recoil angle ($\alpha$).

$$\sigma_\delta = \sqrt{\sigma_{\delta_\text{p}}^2 \cos^2 \alpha + \sigma_{\delta_\text{n}}^2 \sin^2 \alpha} \tag{A1}$$

$\delta_\text{p}$ is calculated from the multiple Coulomb scattering of the recoil protons ($\sigma_\text{scattering}$), and fitting accuracy of the track in the TPC ($\sigma_\text{fit}$).

$$\sigma_{\delta_\text{p}} = \sqrt{\sigma_\text{scattering}^2 + \sigma_\text{fit}^2} \tag{A2}$$

### A.1 $\sigma_\text{scattering}$

The multiple Coulomb scattering is unavoidable in the measurement of proton tracks and affects the accuracy of the reconstructed direction. The scattering angle $\varphi_\text{scattering}$ is defined as the angle between the projection of the final direction and the initial direction in a plane, e.g., the plane including the initial direction and perpendicular to the projection plane $\lambda$ (Figure A1). $\varphi_\text{scattering}$ obeys the Gaussian distribution at small scattering angles and obeys the Rutherford scattering at large angles. The standard deviation of multiple scattering angles of incident ions in a material is defined by the Highland Formula[20]:

$$\varphi_\text{scattering} = \frac{13.6 \text{ MeV}}{\beta c p} z \sqrt{\frac{l_0}{X_0}} \left[1 + 0.038 \ln\left(\frac{l_0}{X_0}\right)\right] \tag{A3}$$

where $p$, $\beta c$, and $z$ are the momentum, velocity, and charge number of the incident particle, respectively, and $l_0$ is the areal density of the material. $X_0$ is the radiation length of the material, and can be calculated by the empirical formula[20]:

$$X_0 = \frac{716.4 \ A}{Z(Z+1) \ln(287 \ Z^{-0.5})} \tag{A4}$$

where $A$ and $Z$ are the mass number and the atomic number of the material. The radiation length of a mixture or a compound is calculated by

$$\frac{1}{X_0} = \sum \frac{w_i}{X_i} \tag{A5}$$

where $X_i$ and $w_i$ are the radiation length and the weight fraction of the $i^\text{th}$ element, respectively.

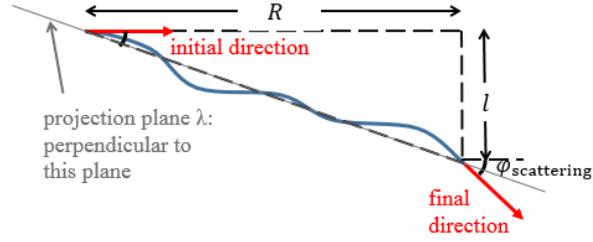

**Figure A1.** Schematic diagram of multiple Coulomb scattering.

As shown in Figure A1, the standard deviation of $\delta_p$ that results from multiple Coulomb scattering is given by[20]:

$$\sigma_{\text{scattering}} = \frac{\sigma_l}{R} = \sigma_{\varphi_{\text{scattering}}} \cdot R \cdot \frac{1}{\sqrt{3}} \cdot \frac{1}{R} = \frac{\sigma_{\varphi_{\text{scattering}}}}{\sqrt{3}} \tag{A6}$$

where $R$ is the thickness of the material, and $l$ is the distance between the final position of the proton and the projection plane.

## A.2 $\sigma_{\text{fit}}$

$\sigma_{\text{fit}}$ is caused by the fitting uncertainty of the recoil proton direction. To simplify the analysis, a coordinate system is set up with the drift direction of the TPC as the $z$-direction and the direction of readout pad rows as the $x$-direction.

As shown in Figure A2, a typical track reconstruction process consists of two steps [21]: (1) Reconstruct the hit positions of each pad row by calculating the weighted average of $y$ and $z$ (the Center of Gravity Method, COG Method); (2) Fit the data points to reconstruct the direction.

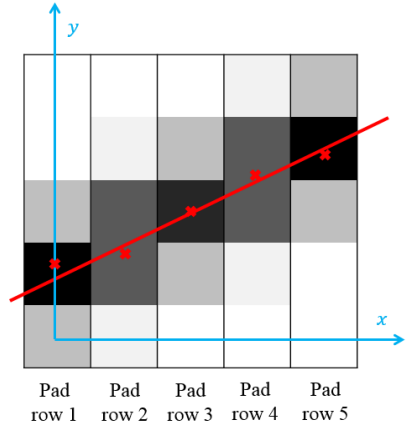

**Figure A2.** Schematic diagram of track reconstruction in the TPC in the $xy$ plane.

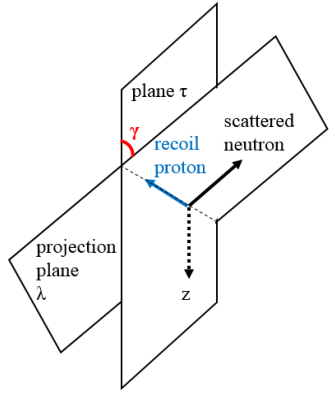

**Figure A3.** Geometric diagram of the projection plane and the plane τ.

Let $\theta_{\text{TPC}}$ and $\varphi_{\text{TPC}}$ be the polar and azimuthal angles, respectively, of the recoil proton direction in this coordinate system. Then, the coordinates of the recoil proton direction are $(\sin\theta_{\text{TPC}}\cos\varphi_{\text{TPC}}, \sin\theta_{\text{TPC}}\sin\varphi_{\text{TPC}}, \cos\theta_{\text{TPC}})$. Let $\tau$ be the plane determined by the direction of the recoil proton and the zenith direction. If $\theta_{\text{TPC}}$ is reconstructed with an error $\mathrm{d}\theta_{\text{TPC}}$, the reconstructed direction deviates from the true direction with an angle $\Delta\theta_0$ in plane $\tau$. The value of $\mathrm{d}\theta_0$ is calculated as

$$\mathrm{d}\theta_0 = \arcsin[(\sin(\theta_{\text{TPC}} + \mathrm{d}\theta_{\text{TPC}})\cos\varphi_{\text{TPC}}, \sin(\theta_{\text{TPC}} + \mathrm{d}\theta_{\text{TPC}})\sin\varphi_{\text{TPC}},$$
$$\cos(\theta_{\text{TPC}} + \mathrm{d}\theta_{\text{TPC}})) \cdot (\sin\theta_{\text{TPC}}\cos\varphi_{\text{TPC}}, \sin\theta_{\text{TPC}}\sin\varphi_{\text{TPC}}, \cos\theta_{\text{TPC}})] = \mathrm{d}\theta_{\text{TPC}} \quad (A7)$$

Similarly, if $\varphi_{\text{TPC}}$ is reconstructed with an error $\mathrm{d}\varphi_{\text{TPC}}$, the reconstructed direction deviates from the true direction with an angle $\mathrm{d}\varphi_0$ perpendicular to plane $\tau$. The value of $\mathrm{d}\varphi_0$ is calculated as

$$\mathrm{d}\varphi_0 = \arcsin[(\sin\theta_{\text{TPC}}\cos(\varphi_{\text{TPC}} + \mathrm{d}\varphi_{\text{TPC}}), \sin\theta_{\text{TPC}}\sin(\varphi_{\text{TPC}} + \mathrm{d}\varphi_{\text{TPC}}), \cos\theta_{\text{TPC}})$$
$$\cdot (\sin\theta_{\text{TPC}}\cos\varphi_{\text{TPC}}, \sin\theta_{\text{TPC}}\sin\varphi_{\text{TPC}}, \cos\theta_{\text{TPC}})] = \mathrm{d}\varphi_{\text{TPC}}\sin\theta_{\text{TPC}} \quad (A8)$$

Let $\vec{v_p}$, $\vec{v_n}$, and $\vec{v_0}$ be the direction of the recoil proton, direction of the scattered neutron, and direction perpendicular to them, respectively. Because the three vectors are perpendicular to each other, they form a set of basis vectors in the three-dimensional space. Let $\vec{v_z}$ be the zenith direction, which is a linear combination of the basis vectors.

$$\vec{v_z} = \sin\varepsilon\cos\gamma\,\vec{v_0} + \sin\varepsilon\sin\gamma\,\vec{v_n} + \cos\varepsilon\,\vec{v_p} \quad (A9)$$

where $\gamma$ and $\varepsilon$ can be calculated from the coordinates of $\vec{v_p}$, $\vec{v_n}$, and $\vec{v_0}$. With an error of $\mathrm{d}\theta_0$ in plane $\tau$, the reconstructed direction of $\vec{v_p}$ is calculated as

$$\vec{v_p}(\mathrm{d}\theta_0) = \cos\mathrm{d}\theta_0\,\vec{v_p} + \sin\mathrm{d}\theta_0\,\frac{\vec{v_z} - (\vec{v_z}\cdot\vec{v_p})\vec{v_p}}{|\vec{v_z} - (\vec{v_z}\cdot\vec{v_p})\vec{v_p}|}$$
$$= \sin\mathrm{d}\theta_0\cos\gamma\,\vec{v_0} + \sin\mathrm{d}\theta_0\sin\gamma\,\vec{v_n} + \cos\mathrm{d}\theta_0\,\vec{v_p} \quad (A10)$$

The reconstruction error of the ARM is

$$\arcsin\{[\vec{v_p}(\mathrm{d}\theta_0) \times \vec{v_n}] \cdot \vec{v_0}\} = \mathrm{d}\theta_0 = \mathrm{d}\theta_{\text{TPC}} \quad (A11)$$

Similarly, with an error of $\mathrm{d}\varphi_0$ perpendicular to plane $\tau$, the reconstructed direction of $\vec{v_p}$ is calculated as

$$\vec{v_p}(\mathrm{d}\varphi_0) = \cos\mathrm{d}\varphi_0\,\vec{v_p} + \sin\mathrm{d}\theta_0\,\frac{\vec{v_p} \times \vec{v_z}}{|\vec{v_p} \times \vec{v_z}|}$$
$$= \sin\mathrm{d}\varphi_0\sin\gamma\,\vec{v_0} + \sin\mathrm{d}\varphi_0\cos\gamma\,\vec{v_n} + \cos\mathrm{d}\varphi_0\,\vec{v_p} \quad (A12)$$

The reconstruction error of the ARM is

$$\arcsin\{[\vec{v_p}(\mathrm{d}\theta_0) \times \vec{v_n}] \cdot \vec{v_0}\} = \mathrm{d}\varphi_0 = \mathrm{d}\varphi_{\text{TPC}}\sin\theta_{\text{TPC}} \quad (A13)$$

Based on Equations (A11) and (A13), $\sigma_{\text{fit}}$ can be calculated as

$$\sigma_{\text{fit}} = \sqrt{\sigma_{\theta_{\text{TPC}}}^2 \cos^2 \gamma + \sigma_{\varphi_{\text{TPC}}}^2 \sin^2 \theta_{\text{TPC}} \sin^2 \gamma} \tag{A14}$$

$\gamma$ varies greatly with the direction of the projection plane $\lambda$, but with the average value of $\cos^2 \gamma$ being 1/2, the average value of $\sigma_{\text{fit}}$ can be estimated as:

$$\overline{\sigma_{\text{fit}}} = \sqrt{\frac{1}{2}\sigma_{\theta_{\text{TPC}}}^2 + \frac{1}{2}\sigma_{\varphi_{\text{TPC}}}^2 \sin^2 \theta_{\text{TPC}}} \tag{A15}$$

$\theta_{\text{TPC}}$ and $\varphi_{\text{TPC}}$ are decided by the fitted slope of $z$ with respect to $x$ ($k_{zx}$) and the slope of $y$ with respect to $x$ ($k_{yx}$). The uncertainties are deduced as:

$$\sigma_{\theta_{\text{TPC}}} = \sqrt{\left(\sin^2 \theta_{\text{TPC}} \cos \varphi_{\text{TPC}} \sigma_{k_{zx}}\right)^2 + \left(\cos^2 \theta_{\text{TPC}} \cos \varphi_{\text{TPC}} \sigma_{k_{yx}}\right)^2} \tag{A16}$$

$$\sigma_{\varphi_{\text{TPC}}} = \cos^2 \varphi_{\text{TPC}} \sigma_{k_{yx}} \tag{A17}$$

If $k_{zx}$ and $k_{yx}$ are fitted using the Least Squares Method, their uncertainties are deduced as[16]:

$$\sigma_{k_{yx}}^2 = \frac{\sum_{i=1}^{n} \frac{1}{\sigma_{y_i}^2}}{\sum_{i=1}^{n} \frac{1}{\sigma_{y_i}^2} \sum_{i=1}^{n} \frac{(iw)^2}{\sigma_{y_i}^2} - \sum_{i=1}^{n} \frac{iw}{\sigma_{y_i}^2} \sum_{i=1}^{n} \frac{iw}{\sigma_{y_i}^2}} = \frac{12}{n(n^2-1)} \frac{\sigma_y^2}{w^2} \tag{A18}$$

$$\sigma_{k_{zx}}^2 = \frac{\sum_{i=1}^{n} \frac{1}{\sigma_{z_i}^2}}{\sum_{i=1}^{n} \frac{1}{\sigma_{z_i}^2} \sum_{i=1}^{n} \frac{(iw)^2}{\sigma_{z_i}^2} - \sum_{i=1}^{n} \frac{iw}{\sigma_{z_i}^2} \sum_{i=1}^{n} \frac{iw}{\sigma_{z_i}^2}} = \frac{12}{n(n^2-1)} \frac{\sigma_z^2}{w^2} \tag{A19}$$

where $w$ is the width of the readout pad of the TPC and $n$ is the number of readout pad rows in the event. If the $y$ and $z$ position of the pad rows are calculated using the COG method, $\sigma_y$ and $\sigma_z$ are the position resolution of the TPC and are calculated as[16][21]:

$$\sigma_y^2 = \frac{1}{N_{\text{eff}}} \left(D_T^2 z + \frac{w^2}{12}\right) + \sigma_{y_0}^2 \tag{A20}$$

$$\sigma_z^2 = \frac{1}{N_{\text{eff}}} \left(D_L^2 z + \frac{w^2}{12 \tan^2 \theta_{\text{TPC}}}\right) + \sigma_{z_0}^2 \tag{A21}$$

where $D_T$ and $D_L$ are the transverse and longitudinal diffusion coefficient of the working gas of the TPC, $\sigma_{y_0}$ and $\sigma_{z_0}$ are the intrinsic position resolution of the detector and electronics system in $y$-direction and $z$-direction, and $N_{\text{eff}}$ is the effective electron number per pad row. As shown in Figure A4, $\sigma_{y_0}$ arises mainly from the deviation of $y$ reconstructed by the COG method at small diffusion width [21]. $N_{\text{eff}}$ can be calculated as[16][21]:

$$N_{\text{eff}} \cong \frac{E_{\text{row}}}{\overline{W}} \left(\frac{1 + \alpha_{\text{Polya}}}{2 + \alpha_{\text{Polya}}}\right) \tag{A22}$$

where $E_{\text{row}}$ is the energy deposition per pad row, $\overline{W}$ is the average ionization energy of the working gas and $\alpha_{\text{Polya}}$ is the factor that determines the shape of the Polya distribution of the GEM module gain.

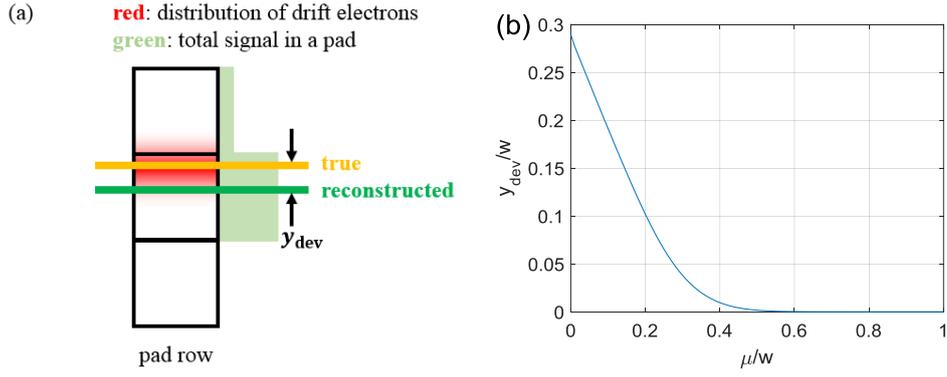

**Figure A4.** Deviation of $y$ constructed by the COG Method.
(a)Geometric diagram; (b) As function of the diffusion width (numerical calculation result, $\mu = D_T\sqrt{z}$).

## A.3 $\sigma_{\delta_n}$

As shown in Figure A5, the coordinate system to be used in this section is set up such that the DOI direction of the scintillation detector is the $z$-direction and the plane determined by the direction of scattered neutron and the $z$-direction is the $xz$ plane. Let $L$ be the distance between the first scattering point and the axial of the scintillation detector, $\theta_{\text{proj}}$ and $\varphi_{\text{proj}}$ be the polar angle and the azimuthal angle of the normal vector of the projection plane ($\vec{v}_{\text{norm}}$) in this coordinate system. The schematic diagrams of some $\theta_{\text{scin}}$ and $\varphi_{\text{scin}}$ are shown in Figure A6. The travel distance $L_{\text{travel}}$ can be calculated as

$$L_{\text{travel}} = \sqrt{L^2 + L^2 \frac{\vec{v}_{\text{norm}} \cdot \vec{x}}{\vec{v}_{\text{norm}} \cdot \vec{z}}} = \sqrt{\frac{1 - \sin^2\theta_{\text{proj}} \sin^2\varphi_{\text{proj}}}{\cos^2\theta_{\text{proj}}}} L \qquad (A23)$$

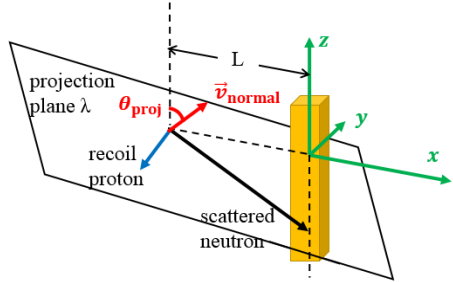

**Figure A5.** Geometric diagram of the projection plane and the plane τ.

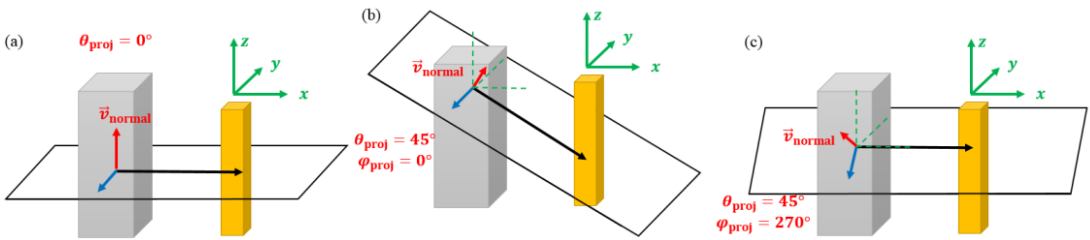

**Figure A6.** The schematic diagrams of some $\theta_{\text{scin}}$ and $\varphi_{\text{scin}}$.

The direction of the scattered neutron is calculated from the position of the two scattering points. The position resolution of the TPC is negligible compared to that of the scintillation detector. $\delta_n$ can be derived from the position uncertainty of the scintillation detector:

$$\sigma_{\delta_n}(\text{DOI}) = \sqrt{\left(\vec{v}_{\text{norm}} \cdot \vec{z} \frac{\sigma_{\text{DOI}}}{L_{travel}}\right)^2 + \left(\sqrt{1-(\vec{v}_{\text{norm}} \cdot \vec{z})^2} \frac{\sigma_d}{L_{travel}}\right)^2}$$

$$= \sqrt{\frac{\cos^2\theta_{\text{proj}}\sin^2\theta_{\text{proj}}}{1-\sin^2\theta_{\text{proj}}\sin^2\varphi_{\text{proj}}}\frac{\sigma_d^2}{L^2} + \frac{\cos^4\theta_{\text{proj}}}{1-\sin^2\theta_{\text{proj}}\sin^2\varphi_{\text{proj}}}\frac{\sigma_{\text{DOI}}^2}{L^2}} \quad (A24)$$

where $\sigma_{\text{DOI}}$ is the position resolution in the $z$-direction, and $\sigma_d$ is the position resolution in $x$-direction and $y$-direction:

$$\sigma_d = \frac{d}{\sqrt{12}} \quad (A25)$$

$\sigma_{\text{DOI}}$ is usually large than $\sigma_d$ in a dual-end readout scintillation detector. With the fact that $\omega$ equals 90°, we can calculate the direction of the scattered neutron without the DOI information. In this case, the direction of the scattered neutron is calculated using the direction of the recoil proton. The resolution of the recoil angle ($\alpha$) is a factor of the angular resolution rather than the DOI resolution.

$$\mathrm{d}\delta_n = \frac{\vec{v}_{\text{norm}} \cdot \vec{z}}{L_{travel}} \frac{L_{travel}\ \mathrm{d}\alpha}{\sqrt{1-(\vec{z} \cdot \vec{v}(\text{proton}))^2}} = \sqrt{\frac{(1-\sin^2\theta_{\text{proj}}\sin^2\varphi_{\text{proj}})}{\sin^2\theta_{\text{proj}}\sin^2\varphi_{\text{proj}}}}\mathrm{d}\alpha \quad (A26)$$

Based on Equations (A24) and (A26), $\delta_n$ is calculated as:

$$\sigma_{\delta_n}(\text{no\_DOI}) = \sqrt{\frac{\cos^2\theta_{\text{proj}}\sin^2\theta_{\text{proj}}}{1-\sin^2\theta_{\text{proj}}\sin^2\varphi_{\text{proj}}}\frac{\sigma_d^2}{L^2} + \frac{(1-\sin^2\theta_{\text{proj}}\sin^2\varphi_{\text{proj}})}{\sin^2\theta_{\text{proj}}\sin^2\varphi_{\text{proj}}}\sigma_\alpha^2} \quad (A27)$$

The uncertainty of the recoil angle ($\sigma_\alpha$) and its average value can be estimated through a similar process with Appendix A.1 and A.2:

$$\sigma_\alpha = \sqrt{\frac{1}{3}\sigma_{\varphi_{\text{scattering}}}^2 + \sigma_{\theta_{\text{TPC}}}^2\sin^2\gamma + \sigma_{\varphi_{\text{TPC}}}^2\sin^2\theta_{\text{TPC}}\cos^2\gamma} \quad (A28)$$

$$\overline{\sigma_\alpha} = \sqrt{\frac{1}{3}\sigma_{\varphi_{\text{scattering}}}^2 + \frac{1}{2}\sigma_{\theta_{\text{TPC}}}^2 + \frac{1}{2}\sigma_{\varphi_{\text{TPC}}}^2\sin^2\theta_{\text{TPC}}} \quad (A29)$$

By comparing Equation (A29) with Equations (A2), (A6), and (A15), it is obvious that $\overline{\sigma_\alpha}$ is equal to $\overline{\sigma_{\delta_p}}$.

## Acknowledgments

We sincerely thank Yina Liu and Hailong Luo of the China Institute of Atomic Energy for useful discussions about the neutron imaging experiment.